\pdfoutput=1
\documentclass[12pt,a4paper,titlepage]{article}
\usepackage[utf8]{inputenc}
\usepackage[top=50pt,bottom=50pt,left=68pt,right=66pt]{geometry}
\usepackage{amsmath}
\usepackage{booktabs}
\usepackage{amssymb}
\usepackage[title]{appendix}
\usepackage{mathrsfs}
\usepackage{float}
\usepackage{multirow}
\usepackage{graphicx,caption,subcaption}
\usepackage[space]{grffile}

\begin{document}
\renewcommand{\arraystretch}{1.3}

\makeatletter
\def\@hangfrom#1{\setbox\@tempboxa\hbox{{#1}}%
      \hangindent 0pt
      \noindent\box\@tempboxa}
\makeatother


\def\un#1{\relax\ifmmode\@@underline#1\else
        $\@@underline{\hbox{#1}}$\relax\fi}


\let\under=\unt                 
\let\ced=\ce                    
\let\du=\du                     
\let\um=\Hu                     
\let\sll=\lp                    
\let\Sll=\Lp                    
\let\slo=\os                    
\let\Slo=\Os                    
\let\tie=\ta                    
\let\br=\ub                     


\def\a{\alpha}
\def\b{\beta}
\def\c{\chi}
\def\d{\delta}
\def\e{\epsilon}
\def\f{\phi}
\def\g{\gamma}
\def\h{\eta}
\def\i{\iota}
\def\j{\psi}
\def\k{\kappa}
\def\l{\lambda}
\def\m{\mu}
\def\n{\nu}
\def\o{\omega}
\def\p{\pi}
\def\q{\theta}
\def\r{\rho}
\def\s{\sigma}
\def\t{\tau}
\def\u{\upsilon}
\def\x{\xi}
\def\z{\zeta}
\def\D{\Delta}
\def\F{\Phi}
\def\G{\Gamma}
\def\J{\Psi}
\def\L{\Lambda}
\def\O{\Omega}
\def\P{\Pi}
\def\Q{\Theta}
\def\S{\Sigma}
\def\U{\Upsilon}
\def\X{\Xi}


\def\ve{\varepsilon}
\def\vf{\varphi}
\def\vr{\varrho}
\def\vs{\varsigma}
\def\vq{\vartheta}


\def\ca{{\cal A}}
\def\cb{{\cal B}}
\def\cc{{\cal C}}
\def\cd{{\cal D}}
\def\ce{{\cal E}}
\def\cf{{\cal F}}
\def\cg{{\cal G}}
\def\ch{{\cal H}}
\def\ci{{\cal I}}
\def\cj{{\cal J}}
\def\ck{{\cal K}}
\def\cl{{\cal L}}
\def\cm{{\cal M}}
\def\cn{{\cal N}}
\def\co{{\cal O}}
\def\cp{{\cal P}}
\def\cq{{\cal Q}}
\def\car{{\cal R}}
\def\cs{{\cal S}}
\def\ct{{\cal T}}
\def\cu{{\cal U}}
\def\cv{{\cal V}}
\def\cw{{\cal W}}
\def\cx{{\cal X}}
\def\cy{{\cal Y}}
\def\cz{{\cal Z}}


\def\Sc#1{{\hbox{\sc #1}}}      
\def\Sf#1{{\hbox{\sf #1}}}      



\def\slpa{\slash{\pa}}                            
\def\slin{\SLLash{\in}}                                   
\def\bo{{\raise-.3ex\hbox{\large$\Box$}}}               
\def\cbo{\Sc [}                                         
\def\pa{\partial}                                       
\def\de{\nabla}                                         
\def\dell{\bigtriangledown}                             
\def\su{\sum}                                           
\def\pr{\prod}                                          
\def\iff{\leftrightarrow}                               
\def\conj{{\hbox{\large *}}}                            
\def\ltap{\raisebox{-.4ex}{\rlap{$\sim$}} \raisebox{.4ex}{$<$}}   
\def\gtap{\raisebox{-.4ex}{\rlap{$\sim$}} \raisebox{.4ex}{$>$}}   
\def\TH{{\raise.2ex\hbox{$\displaystyle \bigodot$}\mskip-4.7mu \llap H \;}}
\def\face{{\raise.2ex\hbox{$\displaystyle \bigodot$}\mskip-2.2mu \llap {$\ddot
        \smile$}}}                                      
\def\dg{\sp\dagger}                                     
\def\ddg{\sp\ddagger}                                   

\font\tenex=cmex10 scaled 1200


\def\sp#1{{}^{#1}}                              
\def\sb#1{{}_{#1}}                              
\def\oldsl#1{\rlap/#1}                          
\def\slash#1{\rlap{\hbox{$\mskip 1 mu /$}}#1}      
\def\Slash#1{\rlap{\hbox{$\mskip 3 mu /$}}#1}      
\def\SLash#1{\rlap{\hbox{$\mskip 4.5 mu /$}}#1}    
\def\SLLash#1{\rlap{\hbox{$\mskip 6 mu /$}}#1}      
\def\PMMM#1{\rlap{\hbox{$\mskip 2 mu | $}}#1}   %
\def\PMM#1{\rlap{\hbox{$\mskip 4 mu ~ \mid $}}#1}       %
\def\Tilde#1{\widetilde{#1}}                    
\def\Hat#1{\widehat{#1}}                        
\def\Bar#1{\overline{#1}}                       
\def\sbar#1{\stackrel{*}{\Bar{#1}}}             
\def\bra#1{\left\langle #1\right|}              
\def\ket#1{\left| #1\right\rangle}              
\def\VEV#1{\left\langle #1\right\rangle}        
\def\abs#1{\left| #1\right|}                    
\def\leftrightarrowfill{$\mathsurround=0pt \mathord\leftarrow \mkern-6mu
        \cleaders\hbox{$\mkern-2mu \mathord- \mkern-2mu$}\hfill
        \mkern-6mu \mathord\rightarrow$}
\def\dvec#1{\vbox{\ialign{##\crcr
        \leftrightarrowfill\crcr\noalign{\kern-1pt\nointerlineskip}
        $\hfil\displaystyle{#1}\hfil$\crcr}}}           
\def\dt#1{{\buildrel {\hbox{\LARGE .}} \over {#1}}}     
\def\dtt#1{{\buildrel \bullet \over {#1}}}              
\def\der#1{{\pa \over \pa {#1}}}                
\def\fder#1{{\d \over \d {#1}}}                 


\def\frac#1#2{{\textstyle{#1\over\vphantom2\smash{\raise.20ex
        \hbox{$\scriptstyle{#2}$}}}}}                   
\def\half{\frac12}                                        
\def\sfrac#1#2{{\vphantom1\smash{\lower.5ex\hbox{\small$#1$}}\over
        \vphantom1\smash{\raise.4ex\hbox{\small$#2$}}}} 
\def\bfrac#1#2{{\vphantom1\smash{\lower.5ex\hbox{$#1$}}\over
        \vphantom1\smash{\raise.3ex\hbox{$#2$}}}}       
\def\afrac#1#2{{\vphantom1\smash{\lower.5ex\hbox{$#1$}}\over#2}}    
\def\partder#1#2{{\partial #1\over\partial #2}}   
\def\parvar#1#2{{\d #1\over \d #2}}               
\def\secder#1#2#3{{\partial^2 #1\over\partial #2 \partial #3}}  
\def\on#1#2{\mathop{\null#2}\limits^{#1}}               
\def\bvec#1{\on\leftarrow{#1}}                  
\def\oover#1{\on\circ{#1}}                              

\def\[{\lfloor{\hskip 0.35pt}\!\!\!\lceil}
\def\]{\rfloor{\hskip 0.35pt}\!\!\!\rceil}
\def\Lag{{\cal L}}
\def\du#1#2{_{#1}{}^{#2}}
\def\ud#1#2{^{#1}{}_{#2}}
\def\dud#1#2#3{_{#1}{}^{#2}{}_{#3}}
\def\udu#1#2#3{^{#1}{}_{#2}{}^{#3}}
\def\calD{{\cal D}}
\def\calM{{\cal M}}

\def\szet{{${\scriptstyle \b}$}}
\def\ulA{{\un A}}
\def\ulM{{\underline M}}
\def\cdm{{\Sc D}_{--}}
\def\cdp{{\Sc D}_{++}}
\def\vTheta{\check\Theta}
\def\fracm#1#2{\hbox{\large{${\frac{{#1}}{{#2}}}$}}}
\def\ha{{\fracmm12}}
\def\tr{{\rm tr}}
\def\Tr{{\rm Tr}}
\def\itrema{$\ddot{\scriptstyle 1}$}
\def\ula{{\underline a}} \def\ulb{{\underline b}} \def\ulc{{\underline c}}
\def\uld{{\underline d}} \def\ule{{\underline e}} \def\ulf{{\underline f}}
\def\ulg{{\underline g}}
\def\items#1{\\ \item{[#1]}}
\def\ul{\underline}
\def\un{\underline}
\def\fracmm#1#2{{{#1}\over{#2}}}
\def\footnotew#1{\footnote{\hsize=6.5in {#1}}}
\def\low#1{{\raise -3pt\hbox{${\hskip 0.75pt}\!_{#1}$}}}

\def\Dot#1{\buildrel{_{_{\hskip 0.01in}\bullet}}\over{#1}}
\def\dt#1{\Dot{#1}}

\def\DDot#1{\buildrel{_{_{\hskip 0.01in}\bullet\bullet}}\over{#1}}
\def\ddt#1{\DDot{#1}}

\def\DDDot#1{\buildrel{_{_{\hskip 0.01in}\bullet\bullet\bullet}}\over{#1}}
\def\dddt#1{\DDDot{#1}}

\def\DDDDot#1{\buildrel{_{_{\hskip 
0.01in}\bullet\bullet\bullet\bullet}}\over{#1}}
\def\ddddt#1{\DDDDot{#1}}

\def\Tilde#1{{\widetilde{#1}}\hskip 0.015in}
\def\Hat#1{\widehat{#1}}


\newskip\humongous \humongous=0pt plus 1000pt minus 1000pt
\def\caja{\mathsurround=0pt}
\def\eqalign#1{\,\vcenter{\openup2\jot \caja
        \ialign{\strut \hfil$\displaystyle{##}$&$
        \displaystyle{{}##}$\hfil\crcr#1\crcr}}\,}
\newif\ifdtup
\def\panorama{\global\dtuptrue \openup2\jot \caja
        \everycr{\noalign{\ifdtup \global\dtupfalse
        \vskip-\lineskiplimit \vskip\normallineskiplimit
        \else \penalty\interdisplaylinepenalty \fi}}}
\def\li#1{\panorama \tabskip=\humongous                         
        \halign to\displaywidth{\hfil$\displaystyle{##}$
        \tabskip=0pt&$\displaystyle{{}##}$\hfil
        \tabskip=\humongous&\llap{$##$}\tabskip=0pt
        \crcr#1\crcr}}
\def\eqalignnotwo#1{\panorama \tabskip=\humongous
        \halign to\displaywidth{\hfil$\displaystyle{##}$
        \tabskip=0pt&$\displaystyle{{}##}$
        \tabskip=0pt&$\displaystyle{{}##}$\hfil
        \tabskip=\humongous&\llap{$##$}\tabskip=0pt
        \crcr#1\crcr}}


\def\eV{\,{\rm eV}}
\def\keV{\,{\rm keV}}
\def\MeV{\,{\rm MeV}}
\def\GeV{\,{\rm GeV}}
\def\TeV{\,{\rm TeV}}
\def\sv{\left<\sigma v\right>}
\def\({\left(}
\def\){\right)}
\def\cm{{\,\rm cm}}
\def\K{{\,\rm K}}
\def\kpc{{\,\rm kpc}}
\def\beq{\begin{equation}}
\def\eeq{\end{equation}}
\def\bea{\begin{eqnarray}}
\def\eea{\end{eqnarray}}


\newcommand{\be}{\begin{equation}}
\newcommand{\ee}{\end{equation}}
\newcommand{\nbe}{\begin{equation*}}
\newcommand{\nee}{\end{equation*}}

\newcommand{\fr}{\frac}
\newcommand{\lb}{\label}

\thispagestyle{empty}

{\hbox to\hsize{
\vbox{\noindent June 2023 \hfill IPMU23-0019 \\
\noindent   revised version \hfill }}}

\noindent
\vskip2.0cm
\begin{center}

{\large\bf Improved model of primordial black hole formation \\ after Starobinsky inflation}

\vglue.3in

Sultan Saburov~${}^{a}$ and Sergei V. Ketov~${}^{a,b,c}$~\footnote{The corresponding author}
\vglue.3in

${}^a$~Interdisciplinary Research Laboratory, Tomsk State University, Tomsk 634050, Russia\\
${}^b$~Department of Physics, Tokyo Metropolitan University, 1-1 Minami-ohsawa, Hachioji, \\
Tokyo 192-0397, Japan \\
${}^c$~Kavli Institute for the Physics and Mathematics of the Universe (WPI),
\\The University of Tokyo Institutes for Advanced Study,  Chiba 277-8583, Japan\\
\vglue.1in

saburov@mail.tsu.ru,  ketov@tmu.ac.jp

\end{center}

\vglue.3in

\begin{center}
{\Large\bf Abstract}  
\end{center}

A new (improved) model of inflation and primordial black hole (PBH) formation is proposed by combining Starobinsky model of inflation, Appleby-Battye-Starobinsky (ABS) model of dark energy and a quantum correction in the modified $F(R)$ gravity.  The energy scale parameter in the ABS model is taken to be close to the inflationary scale, in order to describe double inflation instead of dark energy.  The quantum correction is given by the term quartic in the spacetime scalar curvature $R$ with a negative coefficient $(-\delta)$ in the $F(R)$ function. It is demonstrated that very good agreement (within $1\sigma$) with current measurements of the cosmic microwave background (CMB) radiation can be achieved by choosing  the proper value of $\delta$, thus solving the problem of low values of the tilt of CMB scalar perturbations in  the earlier proposed model in arXiv:2205.00603.  A large (by the factor of $10^7$ against CMB) enhancement in the power spectrum of scalar perturbations is achieved by fine-tuning the parameters of the model. It is found by numerical analysis that it can lead to formation of asteroid-size PBH with the masses up to $10^{20}$ g, which may form dark matter in the current universe.

\newpage

\section{Introduction}

This year is the centennial anniversary of Friedmann's prediction for expanding universe, which was based on his solution to Einstein's equations. Over the recent years the expanding universe was extended by other important features such as cosmological inflation, dark matter and dark energy.

A paradigm of inflation in the early universe, proposed the long time ago \cite{Guth:1980zm,Linde:1981mu} as a possible
solution to internal (flatness, horizon, initial conditions, structure formation) problems in theoretical cosmology, is currently well supported by precision measurements of the cosmic microwave background (CMB) radiation by WMAP and Planck satellite missions combined with recent BICEP/Keck data \cite{Planck:2018jri,BICEP:2021xfz,Tristram:2021tvh}. The original (1980) Starobinsky model of inflation \cite{Starobinsky:1980te} gives a simple theoretical description of inflation by using only gravitational interactions, in very good agreement with the current CMB measurements. It is therefore natural to extend the Starobinsky model of inflation by including more features such as
 production of primordial black holes (PBH). The PBH formation may explain the origin of black holes and dark matter in the current universe
\cite{Novikov:1967tw,Hawking:1971ei,Barrow:1992hq,Ivanov:1994pa,Carr:2020gox,Escriva:2022duf}. 

In modern terms, see e.g., Refs.~\cite{Ketov:2012jt,Ketov:2021fww,Ketov:2022qwj} for more details, the Starobinsky model is the special case of modified gravity theories with the action 
\be \lb{MGaction}
 S = \fracmm{M_{\rm Pl}^2}{2} \int d^4x \sqrt{-g} \,F(R)~~.
\ee
The Starobinsky $F$-function of spacetime scalar curvature $R$ is given by
\begin{equation} \lb{Fstar}
	F_S(R)=R+\fracmm{R^2}{6 M^2}~.
\end{equation}
Here $M_{\rm Pl}=(8\p G)^{-1/2}=2.435\times 10^{18}$ GeV is the reduced Planck mass, and $M$ is the only (mass) parameter. The known CMB amplitude determines $M\approx 1.3\times 10^{-5}M_{\rm Pl}$, so that the Starobinsky model has no free parameters. 

An $F(R)$-gravity model is well known to be equivalent to the quintessence (scalar-tensor) gravity  model in terms of the inflaton scalar field $\phi$ with the scalar potential $V(R(\f))$ in the parametric form, see e.g.,  Refs.~\cite{Maeda:1988ab,Ivanov:2021chn} for a derivation,
\begin{equation}\label{duality}
V(R)=  M_{\rm Pl}^2 \fracmm{F'R-F}{2(F')^2}~,\quad	\phi(R)=\sqrt{\fracmm{3}{2}}M_{\rm Pl}\ln F'~,
	\end{equation}
where the primes denote the derivatives with respect to the given variable $(R)$. It is usually impossible to analytically derive the inverse function $R(\phi)$ from a given function $\f(R)$ with the notable exception in the Starobinsky case (\ref{Fstar}) where one gets a simple and well-known answer,
\be \lb{Vstar}
V_S(\phi) = \fracmm{3}{4} M^2_{\rm Pl}M^2\left[ 1- \exp\left(-\sqrt{\fracmm{2}{3}}\phi/M_{\rm Pl}\right)\right]^2~.
\end{equation}

To form PBH out of gravitational collapse of large (curvature) perturbations in the early universe, one needs a large enhancement of the power spectrum of scalar perturbations by six or seven orders of magnitude against the CMB spectrum. In the context of single-field models of inflation, it can be achieved in the double inflation scenario via engineering a near-inflection point in the inflaton scalar potential at lower (than inflation) scales \cite{Ivanov:1994pa,Garcia-Bellido:2017mdw}.  The potential  (\ref{Vstar}) does not have an inflection point and, hence, does not lead to PBH production. However, one can modify the Starobinsky $F(R)$-gravity function (\ref{Fstar}) by extra terms that lead to an inflection point. It should be done in agreement with CMB observables leading to constraints on eligible $F$-functions. Moreover, there are other conditions such as absence of ghosts and singularities, and the correct Newtonian limit. Once all the necessary conditions are satisfied, one has to achieve the required enhancement of the power spectrum and generate PBH with the masses beyond the Hawking radiation limit of $10^{15}$ g because, otherwise, all those PBH would evaporate before the present times.

These problems were partially solved in Ref.~\cite{Frolovsky:2022ewg} by adding to the $F(R)$-gravity function (\ref{Fstar}) the additional terms known in the literature as Appleby-Battye-Starobinsky (ABS) model of dark energy \cite{Appleby:2009uf}, and described by the $F$-function
\begin{equation} \lb{ABSf}
	F_{ABS}(R)=(1-g)R + gE_{AB} \ln \left[\fracmm{\cosh \left(\frac{R}{E_{AB}}-b\right)}{\cosh (b)}\right]~,
\end{equation}
where the Appleby-Battye parameter $E_{AB}$ has been introduced as
\begin{equation} \lb{ABp}
	E_{AB}=\fracmm{R_0}{2g\ln(1+e^{2b})}
\end{equation}
with the new dimensional scale $R_0$ and the dimensionless positive parameters $g$ and $b$. The function (\ref{ABSf}) was carefully chosen in Ref.~\cite{Appleby:2009uf} in order to meet the no-ghost (stability) conditions in modified $F(R)$ gravity, which are given by
\be \lb{nogh}
F'(R)>0 \quad {\rm and} \quad F''(R) >0~,
\ee 
avoid singularities, obey the Newtonian limit and mimic a positive cosmological constant representing the dark energy for $R\gg R_0$ with proper values of the parameters $g$ and $b$ defining the shape of the scalar potential. To describe the present dark energy in the universe, the parameter $R_0$ representing the vacuum value of the scalar curvature should be very small, $\sqrt{R_0}\sim 10^{-33}$ eV and, hence, the AB-parameter also.

The shape of the $F$-function (\ref{ABSf}) thus leads to a meta-stable de Sitter vacuum after inflation and, hence, 
a near-inflection point  in the potential. In Ref.~\cite{Frolovsky:2022ewg}, the dark energy value of $\sqrt{R_0}$ was replaced by a much higher value of the order $M$  below the inflationary scale $H_{\rm inf.}\sim 10^{14}$ GeV,~\footnote{Accordingly, we changed the notation for the AB-parameter denoted by $\e_{AB}$ in Ref.~\cite{Appleby:2009uf} to $E_{AB}$.} and the new model with the $F$-function  
\be \lb{FKS}
F_{\rm modified}(R) = F_{ABS}(R) + \fracmm{R^2}{6M^2} 
\ee
was shown to lead to the desired enhancement of the power spectrum of scalar perturbations and the formation of PBH with the asteroid-size masses of the order $10^{19}$ g, which may form the whole dark matter in the present universe. In order to get these results, the parameters $g$ and $b$ were fine-tuned but agreement with the observed CMB tilt $n_s$ of scalar perturbations was not good enough (outside $1\s$ but within $3\s$ lower) in Ref.~\cite{Frolovsky:2022ewg}.

In this communication we propose the improved (new) model of Starobinsky inflation and PBH formation, having very good agreement (within $1\s$) to CMB measurements. The new model is defined in Sec.~2 by combining phenomenological and theoretical considerations. Inflation is studied in Sec.~3. The power spectrum is derived in Sec.~4 together with the PBH masses. Sec.~5 is our Conclusion.

\section{The new model}

Our improved modified gravity model is (phenomenologically) defined by the $F$-function
\begin{equation} \lb{Ff}
	F(R)=(1-g_1) R + gE_{AB}  \ln \left[\fracmm{\cosh \left(\frac{R}{E_{AB}}-b\right)}{\cosh (b)}\right]
	+\fracmm{R^2}{6 M^2} - \d\fracmm{R^4}{48M^6}~~,
\end{equation}
where we have changed the coefficient at the first term with another parameter $g_1\neq g$ and have added the new term quartic in $R$ with the new parameter $\d$. The AB parameter $E_{AB}$ is still defined by Eq.~(\ref{ABp}) where 
$R_0=\b M^2$ with yet another parameter $\b$. All the parameters $(g_1,g,b,\b,\d)$ are dimensionless by definition.

The significance of each term in Eq.~(\ref{Ff}) can be explained as follows.

The second term  in Eq.~(\ref{Ff}) becomes approximately linear in $R$ both for small and large values of $R/E_{AB}$, and thus correlates with the first term linear in $R$.  Hence, in the low-curvature regime for small values of $R/E_{AB}$, consistency with gravity measurements inside the Solar system requires the  Einstein-Hilbert effective action, which implies the relation
\be \lb{g1}
g_1= - g \tanh b~.
\ee

Starobinsky inflation is essentially governed by the third term quadratic in $R$ in Eq.~(\ref{Ff}), which leads to inflaton slow-roll (SR) in the 
high-curvature regime for the values of $R/M^2$ between $220$ and $10$ \cite{Ivanov:2021chn}. The scalar potential (\ref{Vstar}) of the Starobinsky model  has the infinite plateau that allows arbitrary duration of inflation, measured by a number $N_e$ of e-folds. However, the
Starobinsky inflation is unstable against quantum gravity corrections of the higher-order in the scalar curvature. As was demonstrated in 
Refs.~\cite{Ketov:2022zhp,Ivanov:2023zgd,Pozdeeva:2023djw}, the leading superstring-inspired quantum correction should be quartic in $R$,
while it eliminates the infinite plateau in the Starobinsky potential and, hence, restricts the maximal number of e-folds. In order to be consistent
with CMB measurements, the value of the $\d$-coefficient should be small enough, for instance, 
$\abs{\d}<3.9\times 10^{-6}$  according to Ref.~\cite{Ketov:2022zhp}. 

It was assumed in Ref.~\cite{Ketov:2022zhp} that the coefficient in front of the $R^4$ term is positive, which led to
the inflaton scalar potential going down and hilltop inflation. In this paper, this coefficient is taken to be negative $(-\d<0)$, which leads to the
potential going up before inflation, see the next Section. The quartic term with $\d>0$ in Eq.~(\ref{Ff}) is also responsible for raising the CMB values of the scalar tilt $n_s$ in the improved model (\ref{Ff}) against those in Ref.~\cite{Frolovsky:2022ewg}, as is demonstrated below. A similar effect was noticed in the modified E-type inflationary models of alpha-attractors and PBH formation, proposed in Ref.~\cite{Frolovsky:2023hqd}.

PBH formation may happen at the energy scales below the inflationary scale, which are governed by the parameter $\sqrt{R_0}$
of the order $M$. Hence, the parameter $\b$ in Eq.~(\ref{Ff}) should be of the order one. The remaining parameters $g$ and $b$ are also of the order one, while their values should be tuned in order to generate a large peak $\sim 10^{-2}$ in the power spectrum of scalar perturbations.
This can only be done numerically by scanning the parameter space as in Refs.~\cite{Frolovsky:2022ewg,Frolovsky:2023hqd}. 

We found that in the model (\ref{Ff})  the parameters $(\b,g,b)$ must be fine-tuned to very specific values, namely, 
\be \lb{parv}
\b\approx 3.00~,\quad g\approx 2.25 \quad {\rm and} \quad b\approx 2.89~,
\ee
because, otherwise, a peak in the power spectrum is either absent or too small or too high.

It is easy to verify that the first and second derivatives of the $F$-function (\ref{Ff}) are positive,
\begin{equation}\label{f1}
	F'(R) = 1+g\tanh(b)+g\tanh \Bigg(\fracmm{R}{E_{AB}}-b\Bigg) +\fracmm{R}{3M^{2}}- \fracmm{\delta R^{3}}{12M^{6}} >0
\end{equation}
and
\begin{equation}\label{f2}
	F''(R) = \fracmm{g}{E_{AB}} {\rm sech}^{2}\Bigg(\fracmm{R}{E_{AB}}-b\Bigg) + \fracmm{1}{3M^{2}} - \fracmm{\delta R^{2}}{4M^{6}} >0~~,
\end{equation}
for the given values of the parameters and the relevant values of $R<240 M^2$.

\section{Inflaton potential and slow-roll}

\begin{figure}[h]
\begin{minipage}[h]{0.5\linewidth}
\center{\includegraphics[width=0.8\linewidth]{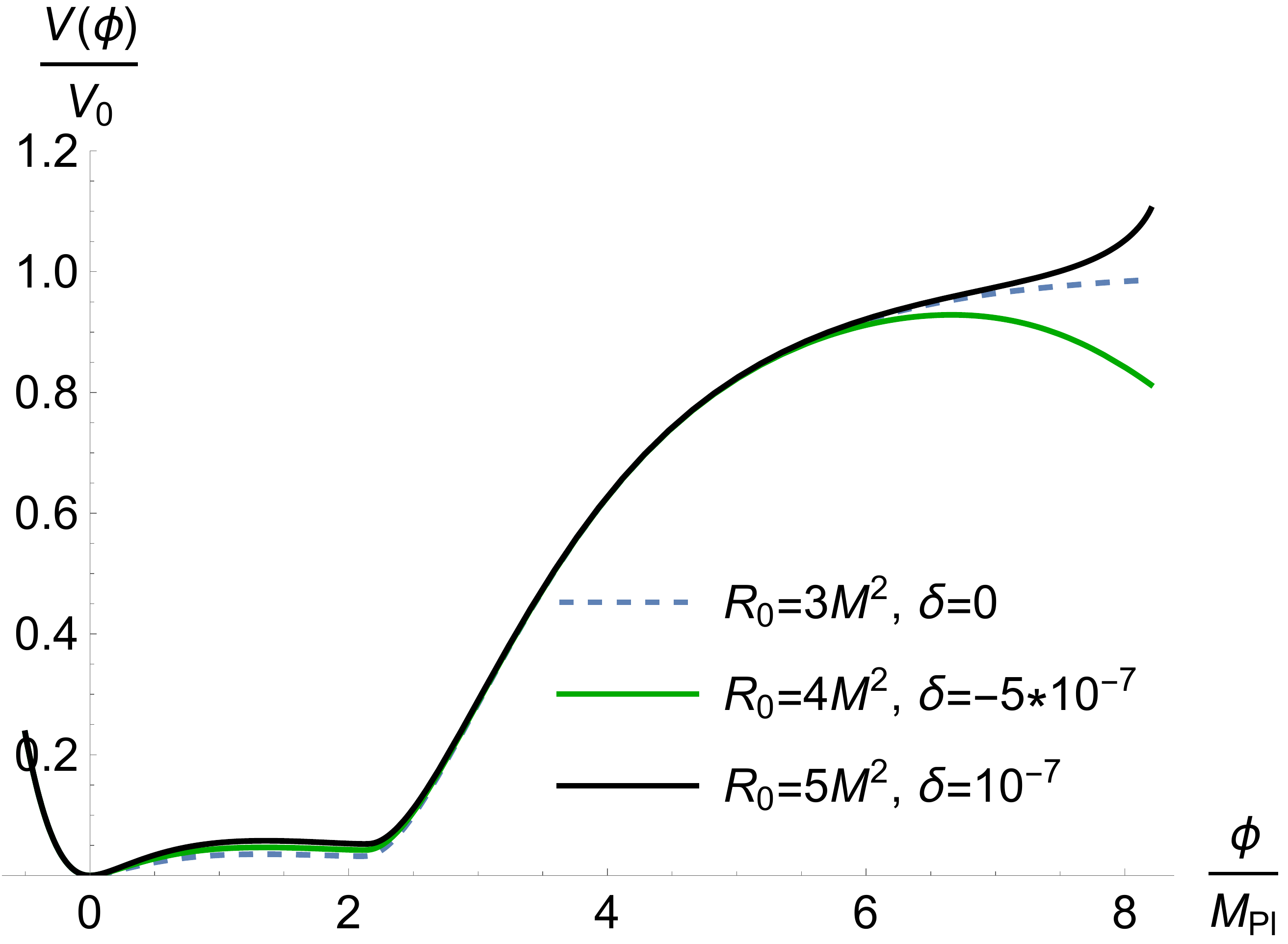} }
\end{minipage}
\hfill
\begin{minipage}[h]{0.5\linewidth}
\center{\includegraphics[width=0.8\linewidth]{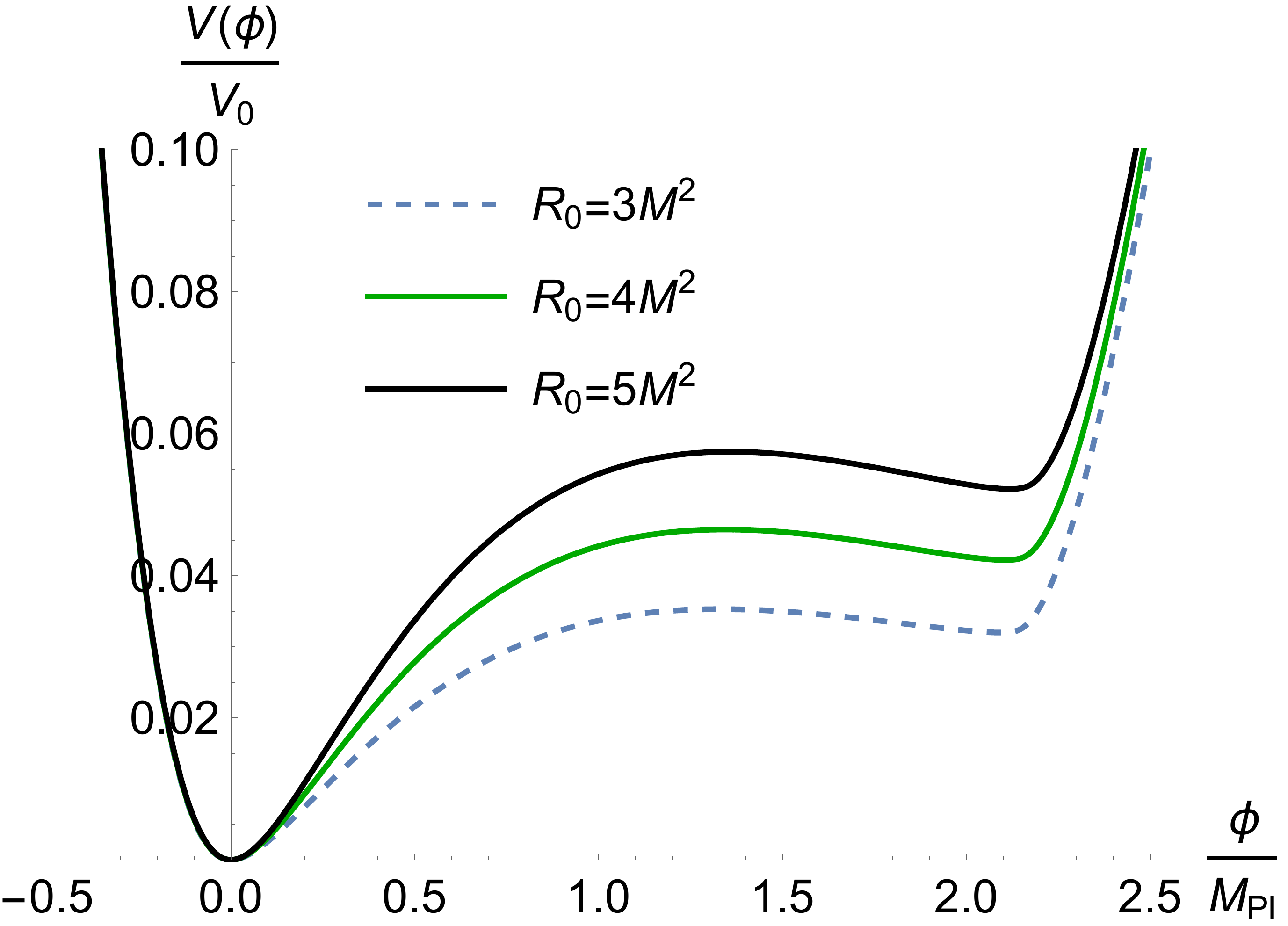} }
\end{minipage}
\caption{\footnotesize The inflaton potential for selected values of $R_0$ and $\d$ at fixed $g=2.25$ and $b=2.89$ with $V_0=\fracmm{3}{4}M^2_{\rm Pl}M^2$ ({\it on the left}). Zooming the potential for low values of $\f/M_{\rm Pl}$ ({\it on the right}).  }
\label{ris:V1}
\end{figure}

According to Eqs.~(\ref{duality}) and (\ref{Ff}), the scalar potential $V(\f)$ of the inflaton field $\f$ in the parametric form (with the parameter $R$) is given by
\begin{equation}\label{pot1}
	\fracmm{V(R)}{M^4_{\rm Pl}} =\frac{1}{2}  e^{-2\sqrt{\frac{2}{3}}\phi/M_{\rm Pl}} \left\{gR\tanh \Bigg(\frac{R}{E_{AB}} -b\Bigg)-gE_{AB} \ln\left[\fracmm{\cosh\left(\frac{R}{E_{AB}}-b\right)}{\cosh(b)}\right] +\fracmm{R^{2}}{6M^{2}} - \fracmm{\delta R^{4}}{16M^{6}} \right\}
\end{equation}
where
\begin{equation}\label{pot2}
\exp\Bigg(\sqrt{\fracmm{2}{3}}\fracmm{\phi}{M_{\rm Pl}}\Bigg) =  1 + g\tanh(b) +g\tanh \Bigg(\frac{R}{E_{AB}}-b\Bigg) + \fracmm{R}{3M^{2}} - \fracmm{\delta R^{3}}{12M^{6}}~.
\end{equation}
The function $\f(R)$ in Eq.~(\ref{pot2}) cannot be inverted in a useful analytic form, so we employ numerical calculations with Mathematica in what follows. The profiles of the inflaton potential $V(\f)$ for selected values of $R_0$ and $\d$ are given in Fig.~\ref{ris:V1}. 

The plot on the left-hand-side of Fig.~\ref{ris:V1} demonstrates that the potential has two plateaus at different scales, it is going up for $\d>0$ before inflation,
while the value of the parameter $\d$ determines the location where the potential goes up at the beginning of SR inflation. The height of the higher (Starobinsky) plateau is determined by $M$, the height of the lower plateau is controlled by $R_0$, the length of the lower plateau is controlled by $g$, and the flatness of the lower plateau is controlled by $b$. The plot on the right-hand-side of Fig.~\ref{ris:V1} shows  a shallow meta-stable de Sitter minimum (dip), a near-inflection point and a small bump (local maximum). 

\begin{figure}[h]
\center{\includegraphics[width=0.5\linewidth]{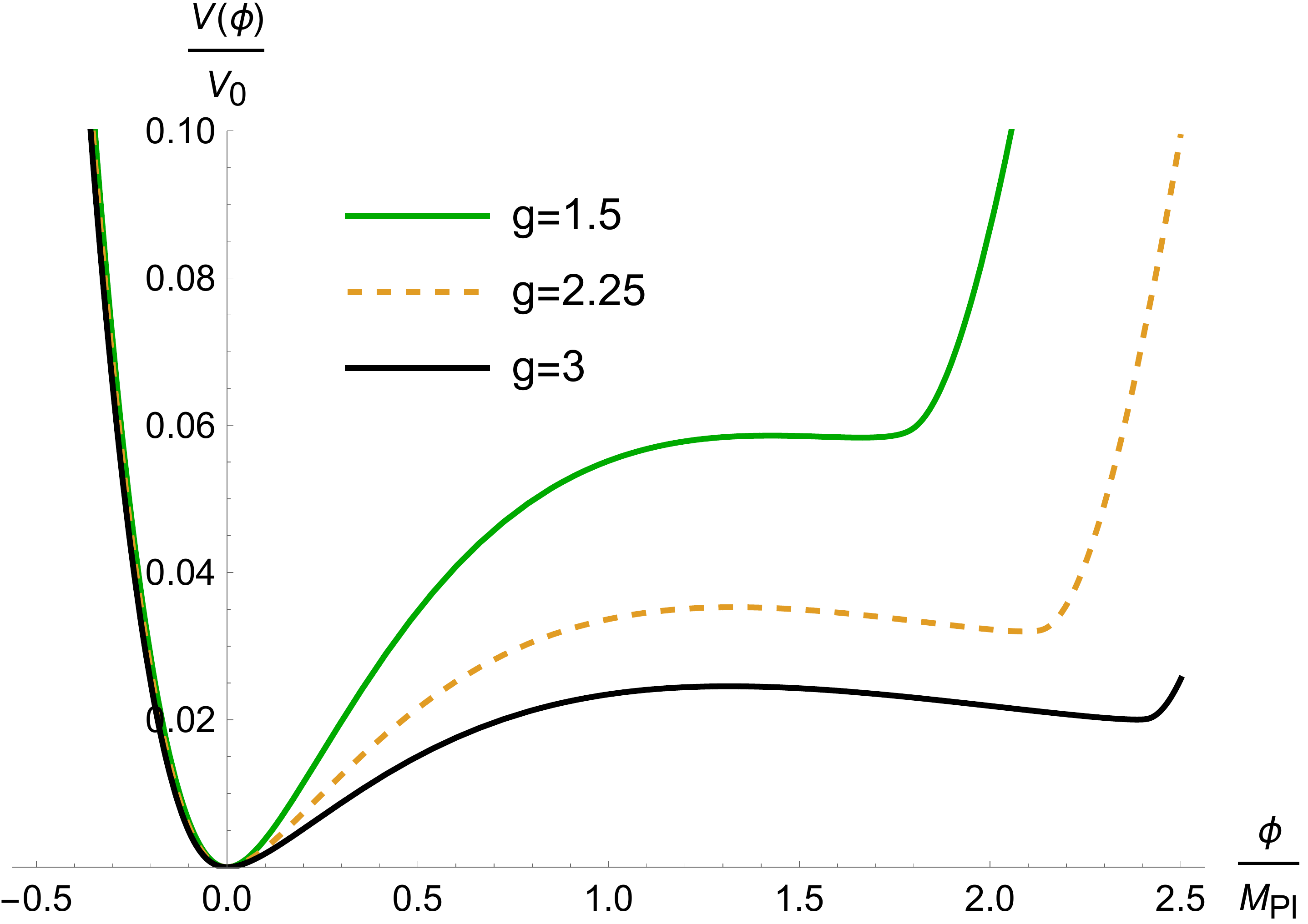}}
\caption{\footnotesize The inflaton potential for selected values of $g$ at fixed $R_0=4M^2$ and $b=2.89$.}
\label{ris:V2}	
\end{figure}

The inflaton potential $V(\f)$ for selected values of $g$ at fixed $R_0=4M^2$ and $b=2.89$ is displayed in Fig.~\ref{ris:V2}. This figure  shows  that duration of the second inflation (length of the lower plateau) is sensitive to the value of the parameter $g$. The longest lower plateau corresponds to $g\approx 3.0$.

\begin{figure}[h]
\begin{minipage}[h]{0.5\linewidth}
\center{\includegraphics[width=0.8\linewidth]{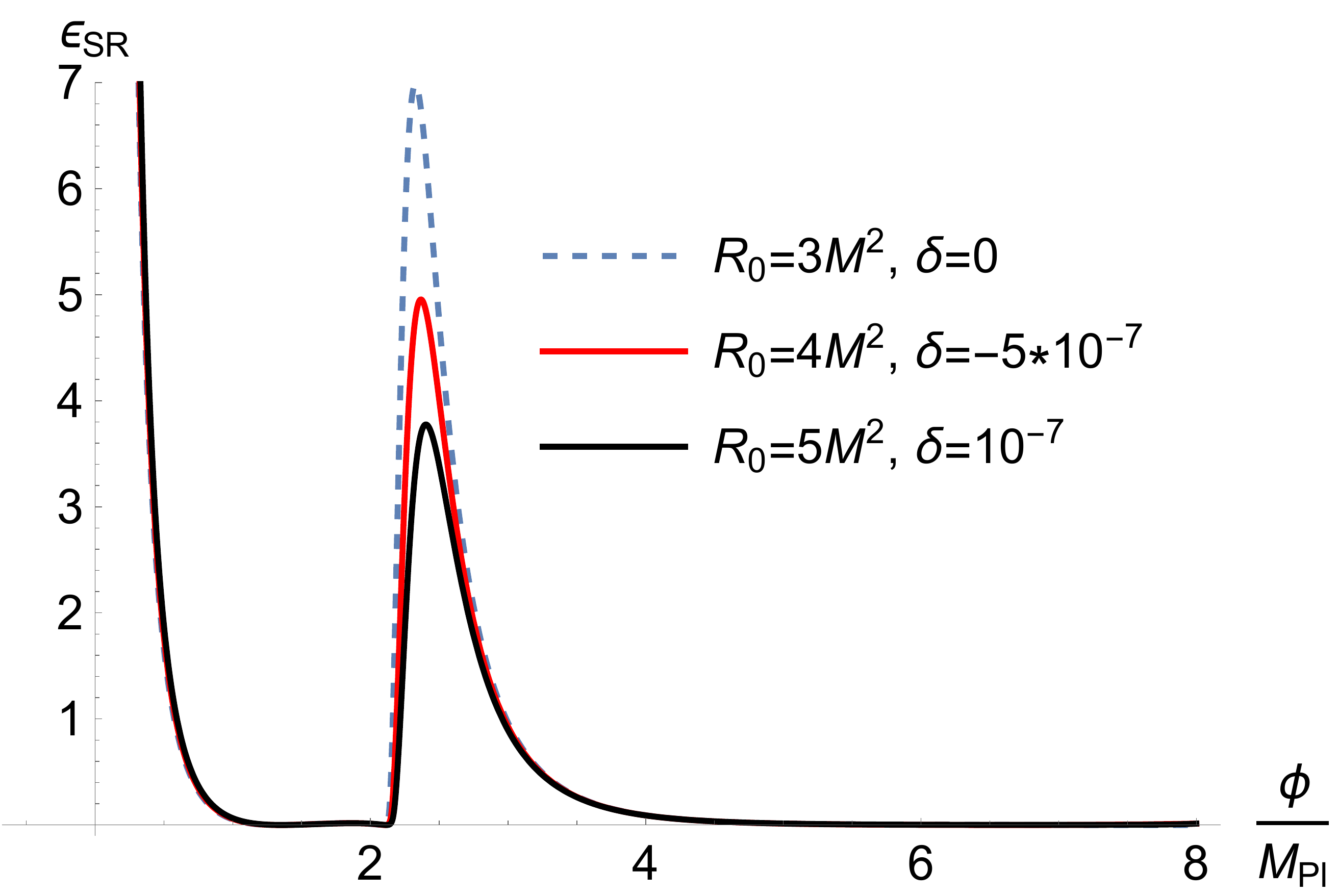} }
\end{minipage}
\hfill
\begin{minipage}[h]{0.5\linewidth}
\center{\includegraphics[width=0.8\linewidth]{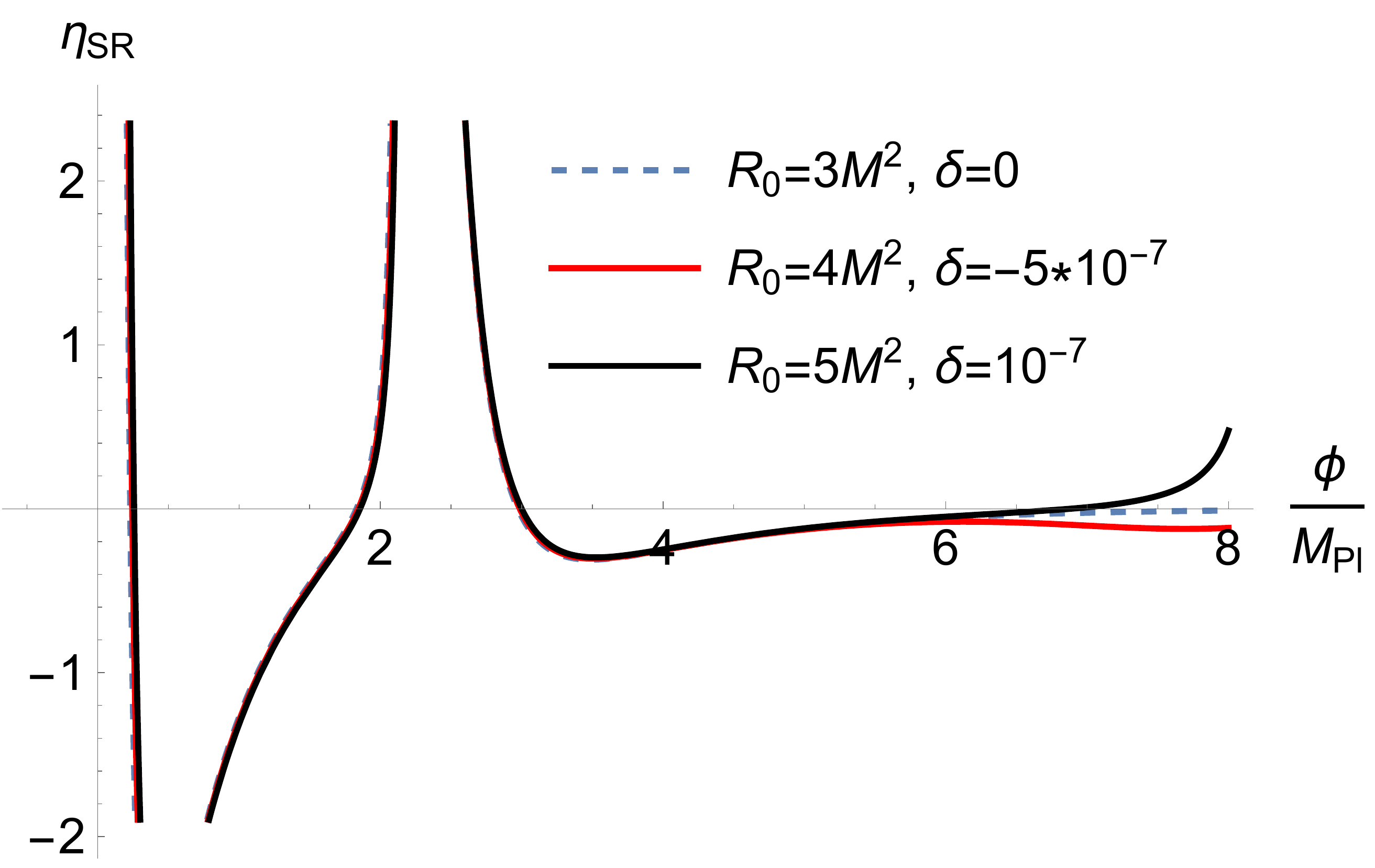} }
\end{minipage}
\caption{\footnotesize The SR parameter $\e_{\rm sr}(\f)$ ({\it on the left}) and the SR parameter $\eta_{\rm sr}(\f)$  ({\it on the right})  for selected values of $R_0$ and $\d$ at fixed $g=2.25$ and $b=2.89$.  The end of Starobinsky inflation  is reached at $\f_{end}\approx 2.98 M_{\rm Pl}$  when $\e_{\rm sr}(\f_{end})\approx 1$.  }
\label{ris:SR}
\end{figure}

The SR conditions are given by smallness of the standard SR parameters, $\e_{\rm sr}\ll 1$ and $\abs{\h_{\rm sr}}\ll 1$, where
\be   \lb{sr} \e_{\rm sr} (\f)= \fracmm{M^2_{\rm Pl}}{2}\left( \fracmm{V'(\f)}{V(\f)}\right)^2\quad {\rm and} \quad 
\h_{\rm sr}(\f) = M^2_{\rm Pl}\fracmm{V''(\f)}{V(\f)}~.
\ee
Duration of inflation is defined by a number $N$ of e-folds,
\be \lb{efolds}
N = \int^{t_{end}}_{t_{in}} H(t) dt \approx \fracmm{1}{M^2_{\rm Pl}}\int ^{\f_{in}}_{\f_{end}} \fracmm{V(\f)}{V'(\f)}d\f~,
\ee
where $H(t)$ is Hubble function. The CMB observable (tilt) $n_s$ of scalar perturbations and the tensor-to-scalar ratio $r$ are related to the values of the SR parameters at the horizon exit with the standard pivot scale  $k_*=0.05~{\rm Mpc}^{-1}$ that (in our model) is close to the scale at the beginning of  SR inflation by $M_{\rm Pl}\delta t\approx 2$ or $\d N\approx 1$ or $\d\f/M_{\rm Pl}\sim 10^{-2}$.  Hence, $\phi_{\rm in}$  can be identified with $\f_{\rm exit}$ at the horizon exit in a derivation of the CMB tilts, leading to 
\be \lb{tilts}
 n_s= 1 + 2\h_{\rm sr}(\f_{\rm in} )- 6 \e_{\rm sr}(\f_{\rm in}) \quad {\rm and} \quad r= 16 \e_{\rm sr}(\f_{\rm in})~.
\ee 
The running SR parameters for selected values of $R_0$ and $\d$ are displayed in Fig.~\ref{ris:SR}. 

It follows from Fig.~\ref{ris:SR} that the peak in $\e_{\rm sr}(\f)$ is sensitive to the value of $R_0$, while the tails of $\h_{\rm sr}(\f)$ are sensitive to the value of $\d$. Therefore, the value of $\d$ affects the value of $n_s$, and then the value of $r$ becomes weakly dependent upon $\d$, in
 agreement with Ref.~\cite{Ketov:2022zhp}.

\begin{figure}[h]
\begin{minipage}[h]{0.5\linewidth}
\center{\includegraphics[width=0.8\linewidth]{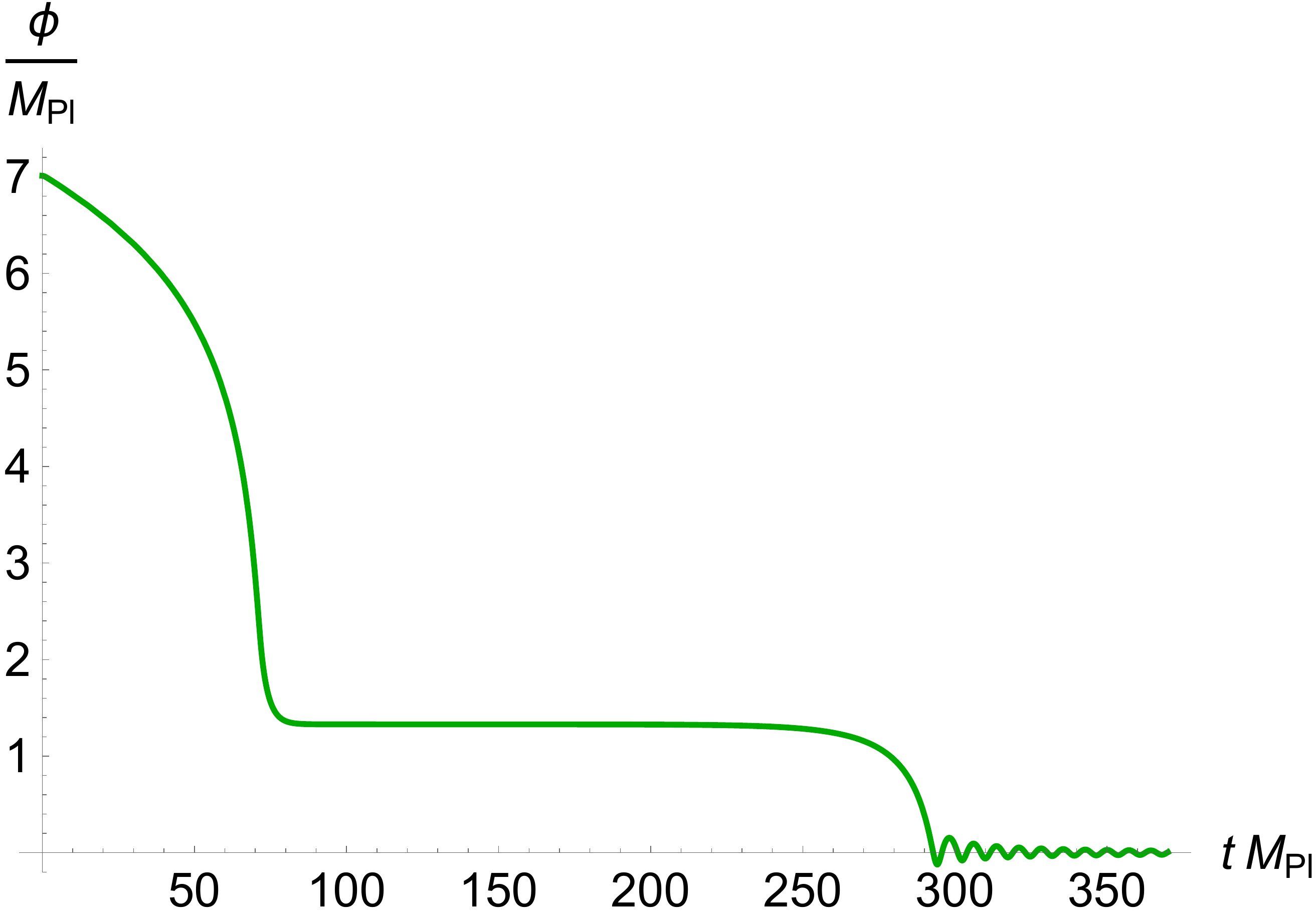} }
\end{minipage}
\hfill
\begin{minipage}[h]{0.5\linewidth}
\center{\includegraphics[width=0.8\linewidth]{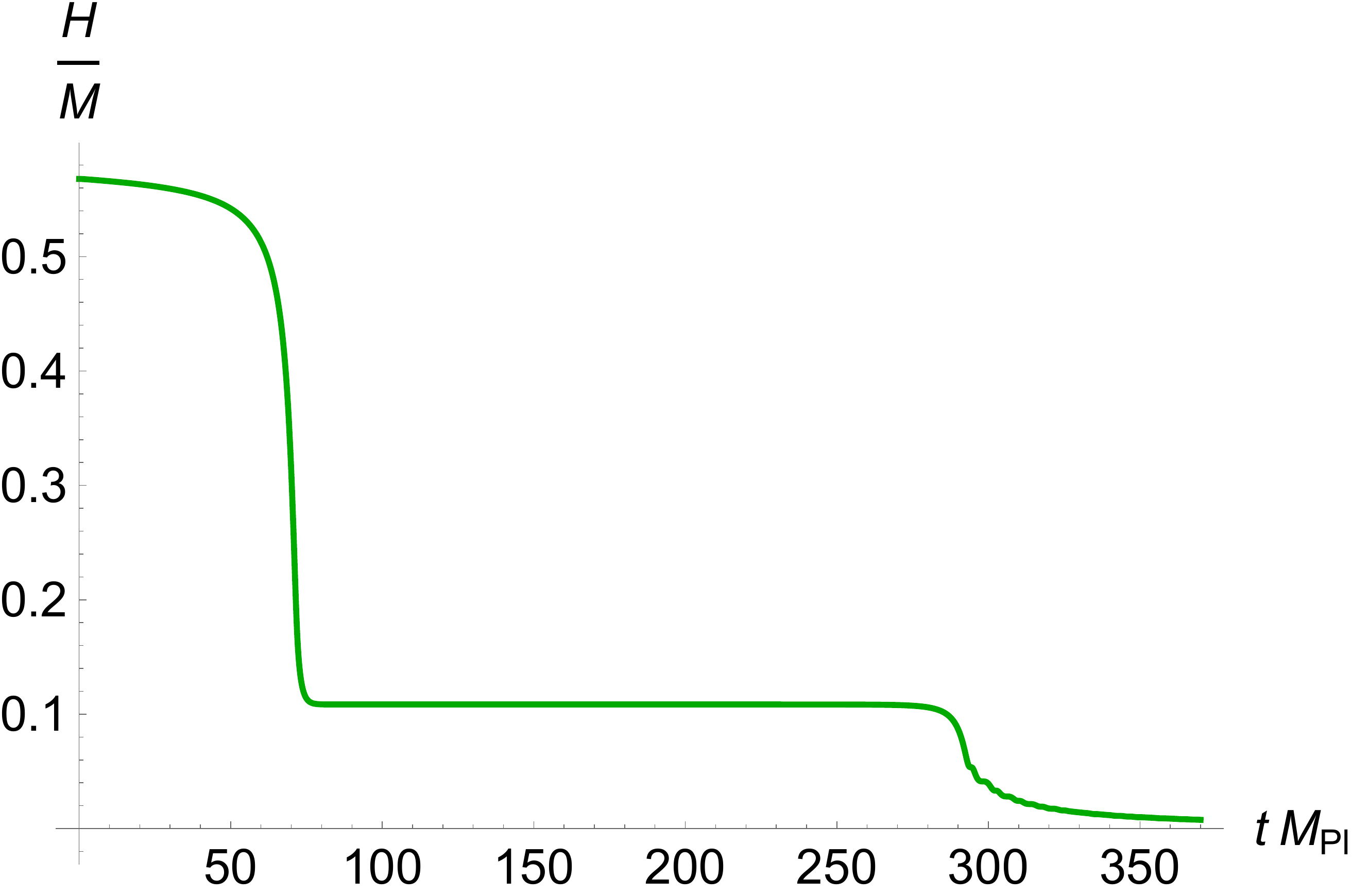} }
\end{minipage}
\caption{\footnotesize The evolution of inflaton field $\f(t)$ and Hubble function $H(t)$ with the initial conditions  $\f_{in}=7.01~M_{\rm Pl}$ and $\dot{\f}_{in}=0$, and the parameters $\d=2.7\cdot 10^{-8}$ and $R_0=3.0~M^2$.}
\label{ris:fH}
\end{figure}

The standard equations of motion of inflaton field $\f(t)$ read 
\begin{equation} \lb{eom}
	  \ddot \phi+3H \dot \phi +V'(\phi)=0, \qquad 
	  3H^2=\fracmm{1}{M^2_{\rm Pl}} \left[ \frac{1}{2} \dot \phi ^2+
	   V(\phi) \right]~,
	  \end{equation}
where the dots denote the derivatives with respect to time $t$. These equations can be numerically solved with given initial conditions 
$\f_{in}$ and $\dot{\f}_{in}$. Due to the attractor-type of Starobinsky inflation the dependence upon initial conditions is weak
\cite{Aoki:2022bvj}.  A solution with the initial conditions $\f_{in}=7.01 M_{\rm Pl}$ and $\dot{\f}_{in}=0$ leading to $\phi_{\rm exit}/M_{\rm Pl}=6.98$ is given in Fig.~\ref{ris:fH} where the Hubble function has two plateaus.

\begin{figure}[h]
\begin{minipage}[h]{0.5\linewidth}
\center{\includegraphics[width=0.8\linewidth]{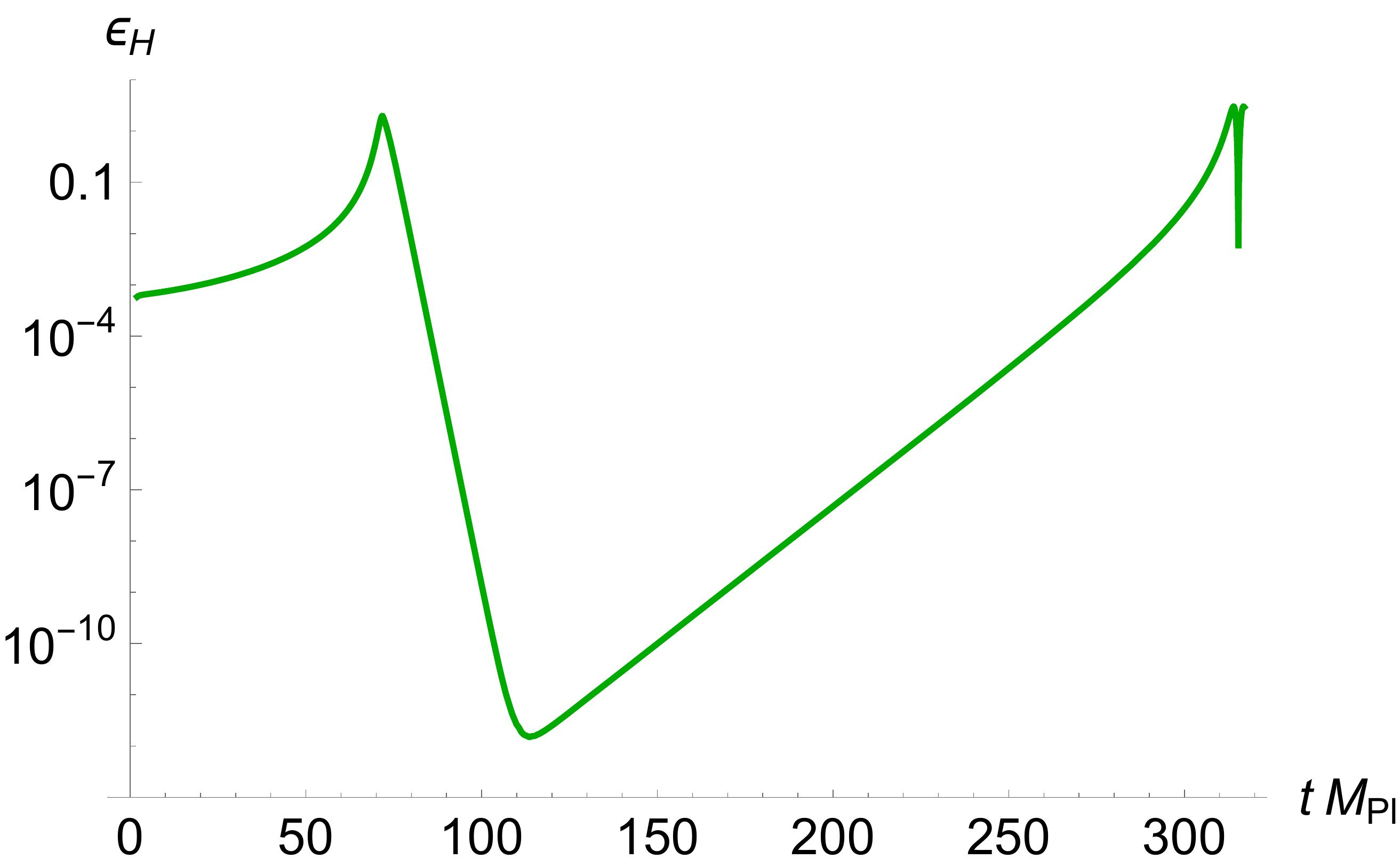} }
\end{minipage}
\hfill
\begin{minipage}[h]{0.5\linewidth}
\center{\includegraphics[width=0.8\linewidth]{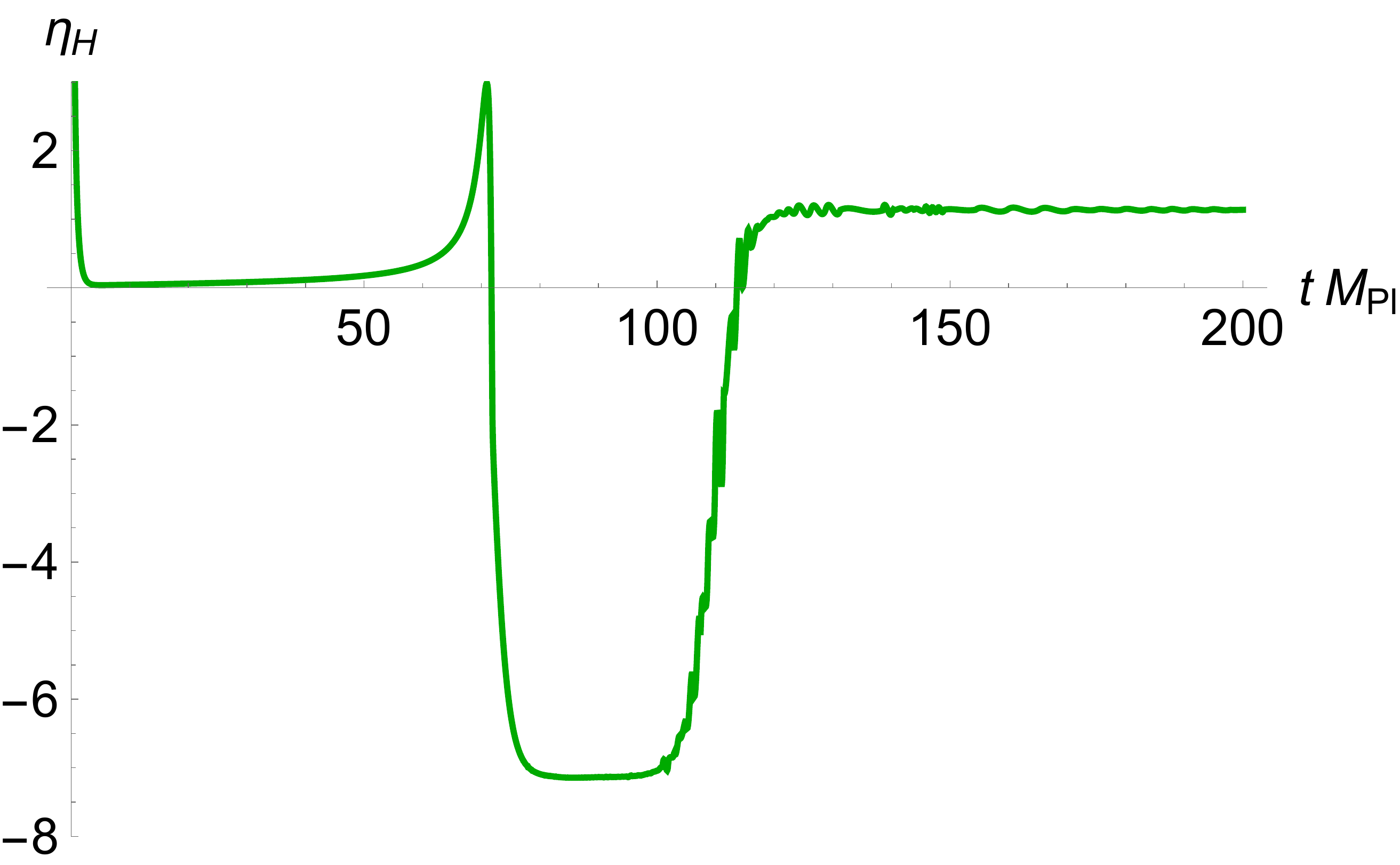} }
\end{minipage}
\caption{\footnotesize The evolution of the Hubble flow parameters $\e_H(t)$ and $\h_H(t)$ with the initial conditions  
$\f_{in}=7.01~M_{\rm Pl}$ and $\dot{\f}_{in}=0$, and the parameters $\d=2.7\cdot 10^{-8}$ and $R_0=3.0~M^2$.}
\label{ris:Hflow}
\end{figure}

Inflaton slowly rolls along both plateaus, whereas between them there is a period of ultra-slow-roll (USR) where dynamics is different
\cite{Germani:2017bcs,Dimopoulos:2017ged}, namely, the acceleration term should be kept but the potential term can be ignored in the first equation of motion (\ref{eom}). Because of that, it is more illuminating to use the Hubble flow parameters outside the CMB region, which are defined by 
\begin{equation}\label{Hflow}
	\epsilon_{H} = -\fracmm{\dot{H}}{H^{2}}~,\quad 
	\eta_{H} = \epsilon_{H} - \fracmm{\dot{\epsilon}_{H}}{2\epsilon_{H} H}~~.
\end{equation}
Their evolution is given in Fig.~\ref{ris:Hflow}. In the USR regime the parameter $\e_H$ becomes very small. The parameter $\h_H$ drops below $-6$, while the corresponding hole (dip) defines the duration of the USR regime.

\section{Power spectrum of scalar perturbations and PBH masses}

\begin{figure}[h]
\center{\includegraphics[width=0.5\linewidth]{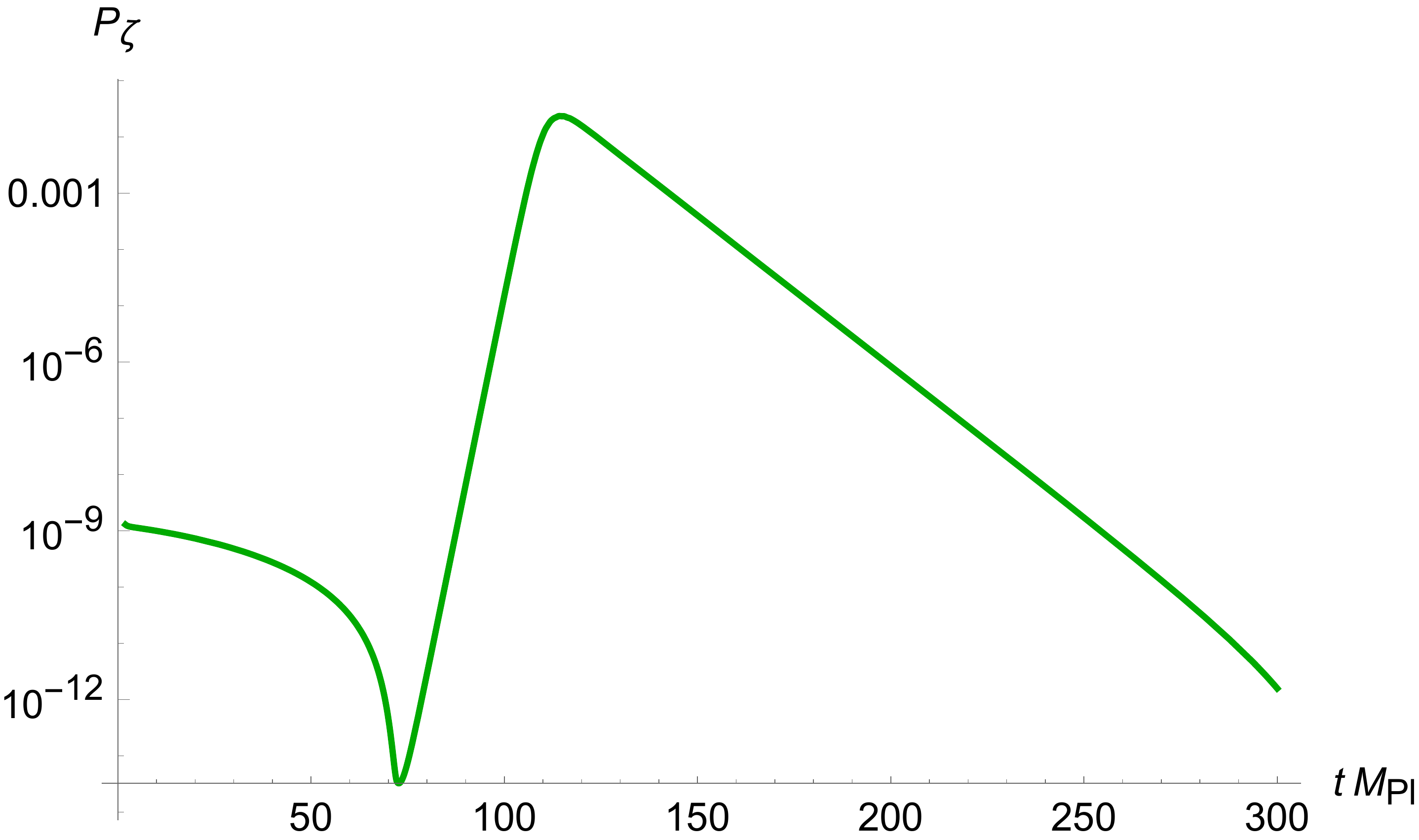}}
\caption{\footnotesize  The primordial power spectrum $P_{\z}(t)$ of scalar (curvature) perturbations, derived from Eq.~(\ref{powersp}). }
\label{ris:P0}	
\end{figure}

The power spectrum $P_{\z}(k)$ of scalar (curvature) perturbations, as a function of scale $k$, is usually derived as a solution to the Mukhanov-Sasaki (MS) equation \cite{Mukhanov:1985rz,Sasaki:1986hm}, which is often possible only numerically.  However, we found that the well-known simple analytic formula 
\cite{Escriva:2022duf,Garcia-Bellido:2017mdw,Karam:2022nym} 
\begin{equation} \lb{powersp}
	P_{\z}(t)=\fracmm{H^2}{8M^2_{\rm Pl}\pi^2\epsilon_H}
\end{equation}
gives a good approximation \cite{Frolovsky:2022ewg,Frolovsky:2023hqd}.  Our new result is given by Fig.~\ref{ris:P0} that shows a large enhancement (peak) in the power spectrum by the factor of $10^7$ against the CMB level (on the left-hand-side from the peak).

The PBH masses can be estimated from the peak in the power spectrum as follows \cite{Pi:2017gih}:
\begin{equation} \lb{pbhm}
M_{\rm PBH}\simeq \fracmm{M_{\rm Pl}^2}{H(t_{\rm peak})} \exp \left[2(N_{\rm total}-N_{\rm peak})+\int_{t_{\rm peak}}^{t_{\rm total} } \epsilon_H(t) H(t) dt    \right]~~,
\end{equation}
where $N_{\rm peak}$ and $t_{\rm peak}$ stand for the peak event, while $N_{\rm total}$ and $t_{\rm total}$ denote the end of the whole inflation comprising three stages (SR/USR/SR).

According to Eq.~(\ref{pbhm}), the PBH masses are exponentially sensitive to the number of e-folds around the inflection point,  $\D N=(N_{\rm total}-N_{\rm peak})$, while the integral in the exponential describes the sub-leading correction that is of the order one.

 Our findings are summarized in Table 1 where the values of the key observables $n_s$, $r$ and $M_{\rm PBH}$ in our model are collected with the  fine-tuned parameters $b=2.89$ and $g=2.25$. The height of the peak is sensitive to $R_0$, whose value $R_0/M^2=3.001$ was chosen to get the height equal to $10^{-2}$.
 
 The tensor-to-scalar ratio $r$ is inside the current observational bound, $r<0.032$, except the first line in the Table (given for comparison only). The tilt $n_s$ of scalar perturbations agrees within $1\s$ with the current CMB  measurements~\cite{Planck:2018jri,BICEP:2021xfz,Tristram:2021tvh}, 
\be \lb{planck}
n_s= 0.9649 \pm 0.0042~.
\ee
\begin{center}
\begin{table}[b]
	\begin{tabular}{|c|c|c|c|c|c|c|c|}
		\hline
$\phi_{in}/M_{\rm Pl}$ & $\d$ & $n_{s}$ & $r$ & $M_{\rm PBH}$, g & $ N_{\rm peak}$ &  $N_{\rm total}$ & $\Delta N$ \\ \hline
6.36 & $2.55\cdot 10^{-7}$ & 0.964959 & 0.0359 & $5.0\cdot 10^{19}$ & 26& 47 & 21\\
6.70 & $8.74\cdot 10^{-8}$ & 0.964905 & 0.0182 & $2.0\cdot 10^{19}$ & 34 & 54 & 20\\
		7.01 & $2.70\cdot 10^{-8}$ & 0.964944 & 0.0095 & $1.0\cdot 10^{20}$ & 43 & 65 & 22\\
		7.07 & $2.05\cdot 10^{-8}$ & 0.964917 & 0.0083 & $2.6\cdot 10^{18}$ & 45 & 64 & 19\\
		7.12  & $1.60\cdot 10^{-8}$ & 0.964925 & 0.0074 & $1.0\cdot 10^{17}$ & 47 & 65 & 18\\
		7.15 & $1.36\cdot 10^{-8}$ & 0.964908& 0.0070& $5.0\cdot 10^{16}$ & 49 & 66 & 17\\
		7.20 & $1.02\cdot 10^{-8}$ & 0.964961 & 0.0062 & $1.6\cdot 10^{15}$  & 51 & 64 & 13\\
		\hline		
	\end{tabular}
	\caption{\footnotesize  CMB observables, PBH masses, and duration of inflation in our model.}
	\end{table}
\end{center}

To get PBH masses beyond the Hawking (black hole) evaporation limit of $10^{15}$ g, so that these PBH can survive in the present universe and form dark matter, the duration $\D N$ should be between $17$ and $22$ e-folds. It also follows from the Table that the total duration of inflation should be 
between $54$ and $66$ e-folds.

Increasing the parameter $\d$ allows us to compensate the decreasing scalar tilt $n_s$. When trying to increase the PBH masses by reducing the total duration of inflation, we find that the value of the tensor-to-scalar ratio $r$ increases and reaches the maximal observationally allowed bound. Increasing $\D N$ even higher is also not possible because it leads to the peak height beyond observational constraints. Therefore, it is not possible to increase the PBH masses beyond the asteroid-size with $10^{20}$ g or, equivalently,  beyond $10^{-13}M_{\rm Sun}$ in our model.

\section{Conclusion}

The main new results of this paper are summarized in the abstract. The modified gravity  framework is entirely based on gravitational interactions, which implies the gravitational origin of both inflation {\it and\/} 
PBH formation in our approach. The good agreement (within $1\sigma$) with CMB observations is achieved by fine-tuning the parameters of the improved model. The PBH masses found are in the mass window that allows the whole dark matter composed of PBH of the asteroid size \cite{Carr:2020gox,Escriva:2022duf}.
The stochastic gravitational waves induced by PBH production in the Starobinsky-like gravity were investigated in Ref.~\cite{Papanikolaou:2021uhe}.

Fine tuning of the parameters in our model is necessary to get the significant enhancement of the power spectrum of scalar perturbations leading to the PBH with masses beyond the Hawking evaporation limit and, hence, the possible PBH dark matter, {\it cf.} Ref.~\cite{Cole:2023wyx}. 

A large peak in the primordial power spectrum may lead to large quantum corrections that may rule out the near-inflection mechanism of PBH production in all single-field models of inflation \cite{Kristiano:2022maq,Kristiano:2023scm,Choudhury:2023rks}. However, validity of the mechanism based on a near-inflection point was 
recently defended in Refs.~\cite{Riotto:2023hoz,Riotto:2023gpm,Firouzjahi:2023ahg,Franciolini:2023lgy,Fumagalli:2023hpa}. Modified gravity offers a different perspective on the issue of quantum corrections when assuming the gravitational origin of inflation (as in the Starobinsky model) and PBH production in the context of $F(R)$ gravity theories, as in our paper. Both  inflation and PBH formation can be destroyed by adding to the $F$-function terms with the higher powers of $R$ describing quantum gravity corrections, with the $R^4$ term being their representative. To avoid that, the coefficients at those terms should be small enough in order to keep validity of the gravitational effective action described by Eq.~(\ref{Ff}). Of course, there should be also a fundamental mechanism in quantum gravity that keeps these coefficients small but this issue is  beyond the scope of this paper.

\section*{Acknowledgements}

SS and SVK were supported by Tomsk State University. SVK was also supported by Tokyo Metropolitan University, the Japanese Society for Promotion of Science under the grant No.~22K03624, and the World Premier International Research Center Initiative (MEXT, Japan).

The authors are grateful to Antonio Iovino Junior, Theodoros Papanikolaou and  Jacopo Fumagalli for correspondence. 

\bibliography{Bibliography}{}
\bibliographystyle{utphys}

\end{document}